\documentclass{ws-p8-50x6-00}

\begin{document}

\title{Polarized Parton Densities in the Nucleon}

\author{Elliot Leader}

\address{Imperial College, University of London,
London WC1E 7HX, England }

\author{Aleksander V. Sidorov}

\address{Bogoliubov Theoretical Laboratory,
JINR, Dubna, Russia}

\author{Dimiter B. Stamenov}

\address{Institute for Nuclear Research and Nuclear Energy,
Sofia, Bulgaria}


\maketitle

\abstracts{
We present a new NLO QCD analysis of the world data on inclusive polarized
deep inelastic scattering. Comparing to our previous analysis: i) the values
of $g_A$ and $a_8=3F-D$
are updated ii)the MRST'99 instead of the MRST'98 parametrization for
the input unpolarized parton densities is used and
iii) the recent SLAC E155 proton data on
the spin asymmetry A1 are included in the analysis. A new set of polarized
parton densities is extracted from the data and the sensitivity of the
results to different positivity constraints is discussed.
}

\section{Introduction}
Deep inelastic scattering (DIS) of leptons on nucleons has remained the
prime source of our understanding of the internal partonic structure of the
nucleon and one of the key areas for the testing of perturbative QCD.
Decades of experiments on unpolarized targets have led to a rather precise
determination of the unpolarized parton densities.
Spurred on by the famous EMC experiment \cite{EMC}
at CERN in 1988, there
has been a huge growth of interest in polarized DIS experiments which yield
more refined information about the partonic structure. Many experiments have
been carried out at SLAC, CERN and DESY.
In this talk we present an updated version of our NLO polarized parton
densities (PD) determined from the world data
\cite{EMC,data,E155p} on inclusive polarized DIS.
Comparing to our previous analysis \cite{spin99}:

~~i) For the axial charges $g_A$ and $a_8$ their updated values are used:
\begin{equation}
g_{A}=\rm {F+D}=1.2670~\pm~0.0035,~~~a_8=3\rm {F-D}=0.585~\pm~0.025~.
\label{ga,3FD}
\end{equation}
~ii) In our ansatz for the input polarized PD
\begin{equation}
\Delta f_i(x,Q^2_0)=A_i~x^{\alpha_i}~f_i^{\rm MRST}(x,Q^2_0)
\label{input}
\end{equation}
we now utilize the MRST'{\bf 99} set \cite{MRST99} of unpolarized parton
densities
$f_i(x,Q^2_0)$ instead of the MRST'{\bf 98} one. In (\ref{input})
$A_i, \alpha_i~$ are free parameters (6 parameters in our fit after using
the sum rules (\ref{ga,3FD}) for the quark polarizations).

iii) The recent SLAC/E155 proton data \cite{E155p} are incorporated in the
analysis.

~iv) The positivity constraints on the polarized PD are discussed.

\section{Method of Analysis}
The spin-dependent structure function of interest, $g^N_1(x,Q^2)$, is a linear
combination of the asymmetries $A^N_{\parallel}$ and $A^N_{\bot}$
(or the related virtual photon-nucleon asymmetries $A^N_{1,2}$) measured
with the target polarized longitudinally or perpendicular
to the lepton beam, respectively. Neglecting as usual the sub-dominant
contributions,
$~A_1^{N}(x,Q^2)~$ can be expressed via the polarized
structure function $~g_1^{N}(x,Q^2)~$ as
\begin{equation}
A_1^{N}(x,Q^2)\cong (1+\gamma^2){g_1^{N}(x,Q^2)\over F_1^{N}(x,Q^2)}~,
\label{assym}
\end{equation}
where  $~F_1^N~$ is the unpolarized structure function and $\gamma^2$ is
a kinematic factor.

All details of our approach to the fit of the data are given in \cite{spin98}.
Here we would like to emphasize only that in our approach
the NLO QCD predictions have been confronted to the data on the spin
asymmetry $A_1^{N}(x,Q^2)$, rather than on the $g_1^{N}(x,Q^2)$.
The choice of $A^N_1$ should minimize the higher twist
contributions which are expected to partly cancel in the ratio
(\ref{assym}), allowing use of data at lower $Q^2$ (in polarized DIS most of
the small $x$ experimental data points are at low $~Q^2~$).
Indeed, we have found \cite{LomConf} that if for $g_1$ and $F_1$ {\it
leading-twist} QCD expressions are used, the HT corrections $h(x)$ to
$A_1(x,Q^2)=A_1^{QCD}(x,Q^2)+h(x)/Q^2$ are negligible and consistent
with zero (see Fig. 1(a)). On other hand,
it was shown \cite{GRSV2000} that if $F_2$ and $R$ (instead of $F_1$ in Eq.
(\ref{assym})) are taken from experiment (as has been done in many
analyses) the higher twist corrections to $A_1$ are sizeable and important.

\section{Results}
The results of analysis are presented in the JET scheme \cite{JET}.
A remarkable property of the JET [and Adler-Bardeen (AB)] schemes is
that $~\Delta \Sigma(Q^2)$, as well as $\Delta s(Q^2)$, are
${\it Q^2}$ {\it independent} quantities. Then, in these schemes
it is meaningful to directly interpret $\Delta \Sigma$ as the
contribution of the quark spins to the nucleon spin and to
compare its value obtained from DIS region with the
predictions of the different (constituent, chiral, etc.) quark
models at low $Q^2$.

As in our previous analysis a very good description of the world
data on $~A_1^N~$ and $~g_1^N~$ is achieved (for the best fit
$\chi^2=155.9$ for 179 DOF). The
\begin{figure}[t]
\epsfxsize=8.9pc 
\epsfbox{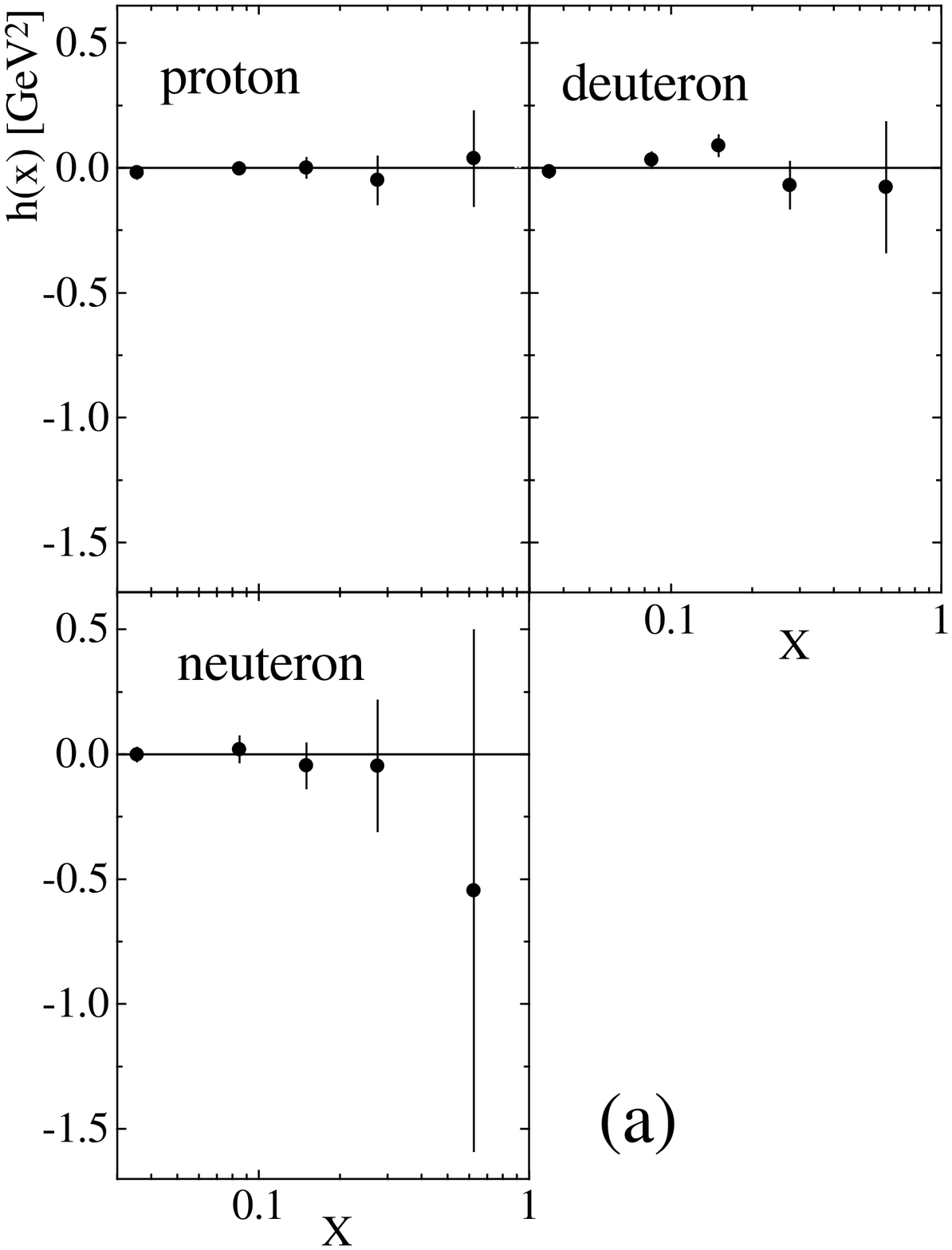} 
\epsfxsize=8.9pc 
\epsfbox{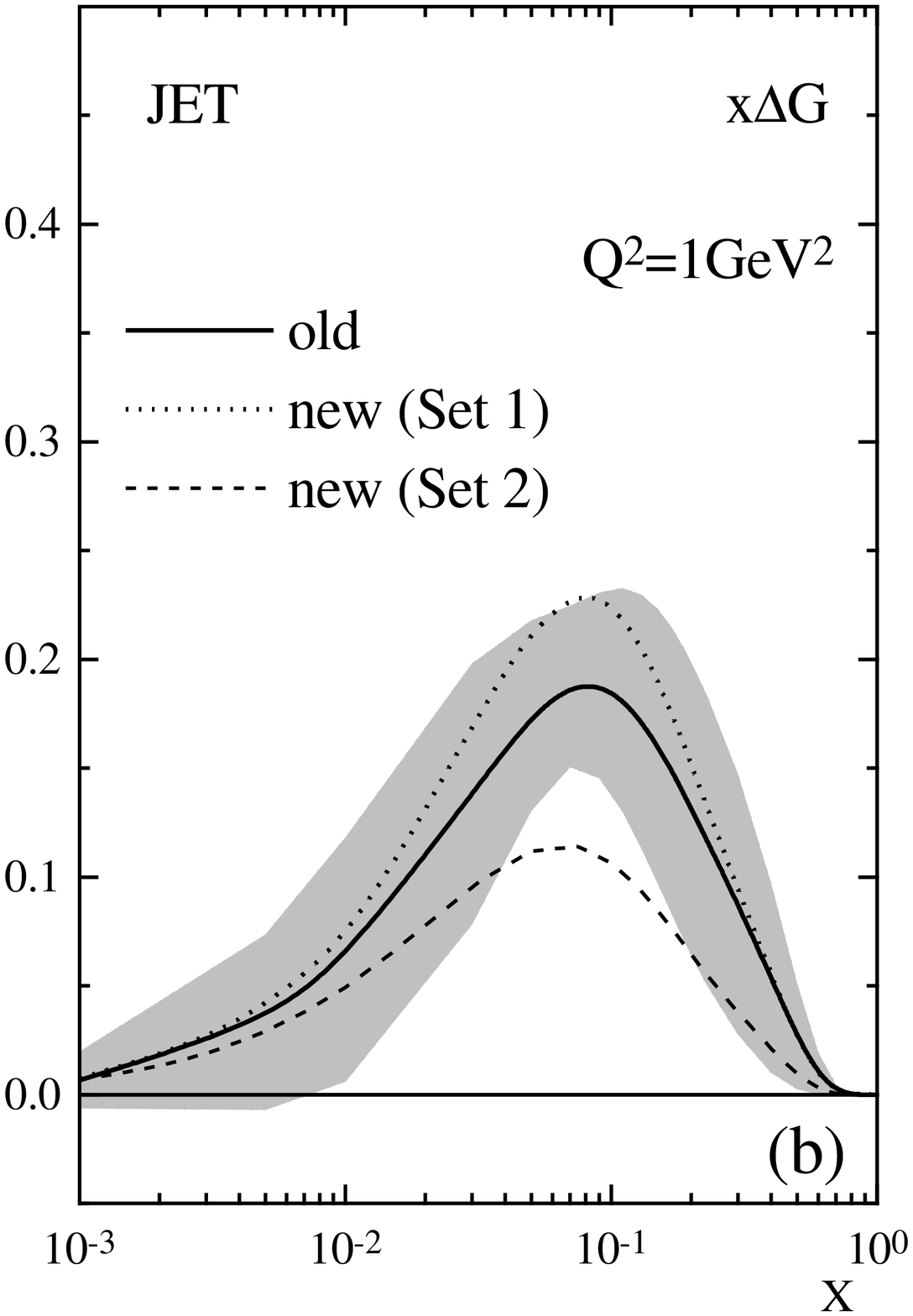} 
\epsfxsize=8.9pc 
\epsfbox{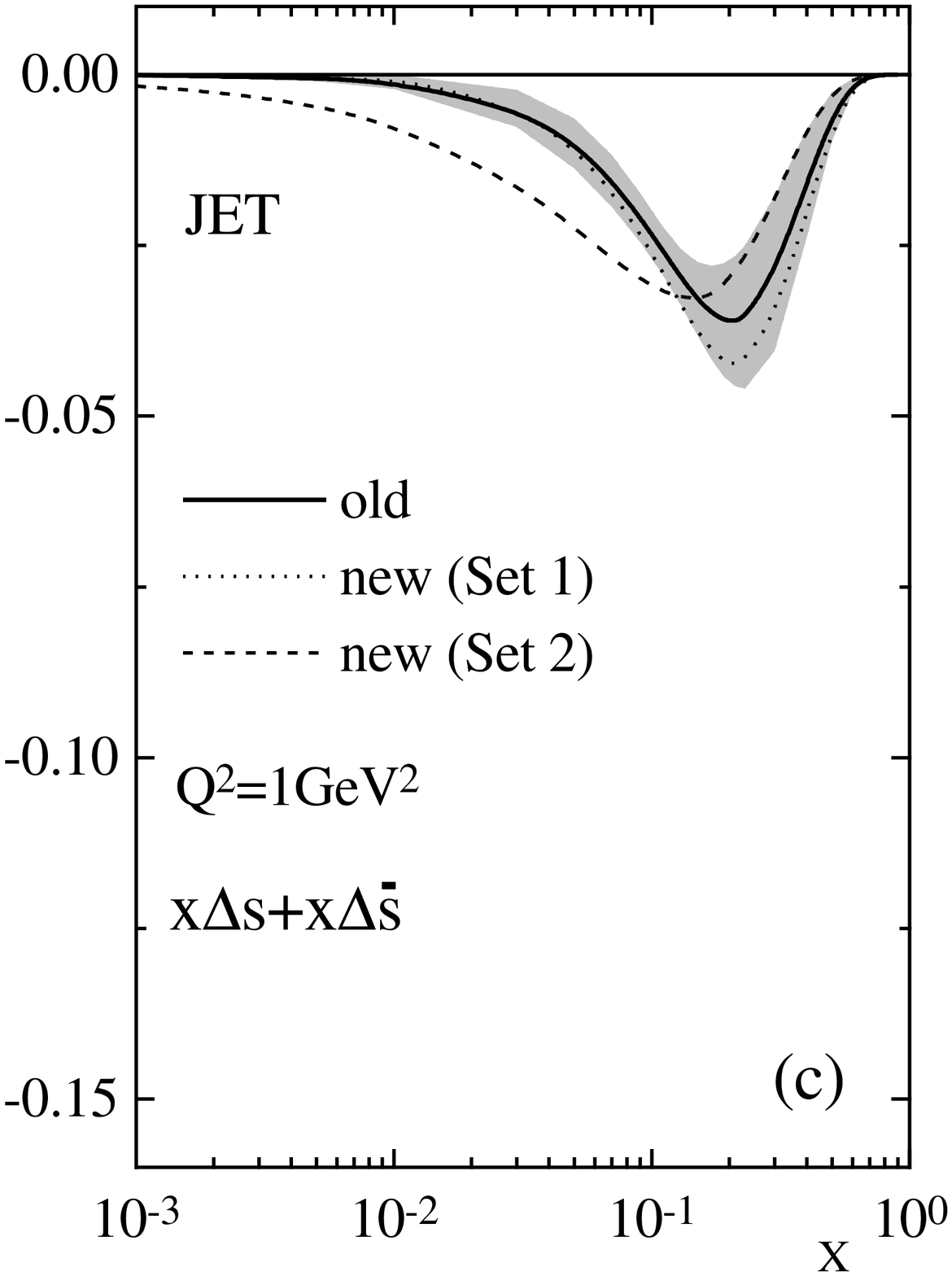} 
\vspace{-2.5mm}
\caption{
 Twist-4 contributions $h^N(x)$ to the spin asymmetry
$A_1^N(x,Q^2)$ extracted from the data (a). The polarized NLO
gluon (b) and strange quark sea (c) densities at $Q^2=1~GeV^2$.
The error bands account for the statistical and systematic
uncertainties.
\label{fig:htdgds}}
\end{figure}
new theoretical curves
for $A_1$ and $g_1$ corresponding to the best fit practically coincide
with the old ones. However, if for the polarized PD the Set2 (see below)
is used the values of $A_1$ at large x increase, in a better
agreement with the large x E155/p data.

We have determined from the data two sets of polarized PD: i) Set1 -
without any constraints on the polarized PD during the fit, ii) Set2 -
fit to the data with imposition of LO positivity constraints determined
by the MRST unpolarized densities. All polarized PD of Set1, with the
exception of the
strange sea density $\Delta s(x)$, are compatible with the LO positivity
bounds imposed by
the MRST unpolarized PD. However, if one uses the {\it more accurate}
LO positivity bounds on $\Delta s(x)$ obtained by using the unpolarized
strange sea density $s(x)_{BPZ}$ (Barone et al. \cite{Barone}),
the Set1 $\Delta s(x)$ lies in the allowed region.
It is important to mention that $s(x)_{BPZ}$ is determined with a higher
accuracy compared to other global fits. The Set1 PD are found to be
within the error bands of the old PD, while the changes of the Set2
$(\Delta s+\Delta\bar{s})(x)$ and $\Delta G(x)$ are larger,
especially for the strange sea quarks (see Fig. 1(b) and Fig. 1(c)).

Finally, let us summarize the changes in the quark and gluon polarizations
at $Q^2=1~GeV^2$ which arise from the new analysis:
\begin{itemize}
\item
\vspace{-1.1mm}
${\bf(\Delta u +\Delta\bar{u})}$ and ${\bf(\Delta
d+\Delta\bar{d})}$ practically do not change.
\item
\vspace{-1.1mm}
the magnitude of the $n=1$ moment, 
${\bf  \vert {\Delta s + \Delta\bar{s}}\vert}$, has increased a little:\\
$0.06 \pm 0.01~(\rm {\bf old})\rightarrow 0.07 \pm 0.01~(\rm {\bf Set1})
\rightarrow 0.09 \pm 0.02~(\rm {\bf Set2})$
\item
\vspace{-1.1mm}
The central value of ${\bf  \Delta \Sigma}$ has decreased:\\
$0.40 \pm 0.04~(\rm {\bf old})\rightarrow 0.37 \pm 0.04~(\rm {\bf Set1})
\rightarrow 0.32 \pm 0.05~(\rm {\bf Set2})$
\item
\vspace{-1.1mm}
The axial charge ${\bf a_0(Q^2)}=\Delta \Sigma_{\rm \overline{MS}}(Q^2)$
is also decreasing:\\
$0.26 \pm 0.05~(\rm {\bf old})\rightarrow 0.21 \pm 0.06~(\rm {\bf Set1,2})$
\item
\vspace{-1.1mm}
For the gluon polarization ${\bf \Delta G}$ we have found that
the positive values of $\Delta G$ lie in the wide range
[0, 1.5] if one takes into account the correlations and the
sensitivity to the SU(3) flavour symmetry breaking.
A  negative $\Delta G$  is still {\it not} excluded from the present
DIS inclusive data.
\end{itemize}

\section*{Acknowledgments}
One of us (D.S.) thanks the Organizers for the financial support of his 
participation in the Meeting. This research was supported by the UK Royal 
Society and the JINR-Bulgaria Collaborative Grants, by the RFBR 
(No 00-02-16696), INTAS 2000 (No 587) and by the Bulgarian National 
Science Foundation.

\end{document}